%% file: EXO-11-098_temp.tex
\pdfoutput=1

\documentclass[11pt,twoside,a4paper,cmspaper,final,collab]{cms-tdr}
\def\svnVersion{175945}\def\cmsCernNo{2012-251}\def\cmsCernDate{\today}\def\cmsMessage{Submitted to Physical Review D}

\begin{document}\cmsNoteHeader{EXO-11-098}

\hyphenation{had-ron-i-za-tion}
\hyphenation{cal-or-i-me-ter}
\hyphenation{de-vices}

\RCS$Revision: 151065 $
\RCS$HeadURL: svn+ssh://svn.cern.ch/reps/tdr2/papers/EXO-11-098/trunk/EXO-11-098.tex $
\RCS$Id: EXO-11-098.tex 151065 2012-10-07 20:50:18Z pvmulder $
\newlength\cmsFigWidth
\ifthenelse{\boolean{cms@external}}{\setlength\cmsFigWidth{0.85\columnwidth}}{\setlength\cmsFigWidth{0.4\textwidth}}
\ifthenelse{\boolean{cms@external}}{\providecommand{\cmsLeft}{top}}{\providecommand{\cmsLeft}{left}}
\ifthenelse{\boolean{cms@external}}{\providecommand{\cmsRight}{bottom}}{\providecommand{\cmsRight}{right}}
\cmsNoteHeader{EXO-11-098} 
\title{Combined search for the quarks of a sequential fourth generation}

\date{\today}

\abstract{
Results are presented from a search for a fourth generation of quarks produced singly or in pairs in a data set corresponding to an integrated luminosity of 5\fbinv
recorded by the CMS experiment at the LHC in 2011. A novel strategy has been developed for a combined search for quarks of the up- and down-type in decay channels
with at least one isolated muon or electron. Limits on the mass of the fourth-generation quarks and the relevant CKM matrix elements are derived in the context of a
simple extension of the standard model with a sequential fourth generation of fermions. The existence of mass-degenerate fourth-generation quarks with masses below
685\GeV is excluded at 95\% confidence level for minimal off-diagonal mixing between the third- and the fourth-generation quarks. With a mass difference of 25\GeV
between the quark masses, the obtained limit on the masses of the fourth-generation quarks shifts by about ${\pm}20\GeV$. These results significantly reduce the allowed parameter space for a fourth generation of fermions.
}

\hypersetup{%
pdfauthor={CMS Collaboration},%
pdftitle={Combined Search for the Quarks of a Sequential Fourth Generation},%
pdfsubject={CMS},%
pdfkeywords={CMS, physics, exotica, toplikeBSM, fourth generation quarks}}

\maketitle 

\section{Introduction} 
\label{sect:intro}

The existence of three generations of fermions has been firmly established experimentally~\cite{Neutrinos}. The possibility of a fourth generation of fermions has not
been excluded, although it is strongly constrained by precision measurements of electroweak observables. These observables are mainly influenced by the mass
differences between the fourth-generation leptons or quarks. In particular, scenarios with a mass difference between the fourth-generation quarks smaller than the mass of
the \PW\ boson are preferred and even fourth-generation quarks with degenerate masses are allowed~\cite{Flacher:2008zq,Buchkremer:2012yy}.

A new generation of fermions requires not only the existence of two additional quarks and two additional leptons, but also an extension of the Cabibbo-Kobayashi-Maskawa (CKM)~\cite{Cabibbo:1963yz,Kobayashi:1973fv} and Pontecorvo-Maki-Nakagawa-Sakata (PMNS)~\cite{Pontecorvo:1957cp,Maki:1962mu} matrices. New CKM (quark mixing) and PMNS (lepton mixing) matrix elements are constrained by the requirement of consistency with electroweak precision measurements~\cite{Asilar:2011bn}.

Previous searches at hadron colliders have considered either pair production \emph{or} single production of \emph{one} of the fourth-generation
quarks~\cite{ATLAS:2012aw,Aad:2012bt,Aad:2012bb,Aad:2012xc,CMS:2012ab,Chatrchyan:2012yea,Chatrchyan:exo099}. The most stringent limits exclude the
existence of a down-type (up-type) fourth-generation quark with a mass below 611 (570)\GeV~\cite{Chatrchyan:2012yea,Chatrchyan:exo099}.
These limits on the quark mass values enter a region where the coupling of fourth-generation quarks to the Higgs field becomes large and perturbative calculations for the weak interaction
start to fail, assuming the absence of other phenomena beyond the standard model~\cite{Chanowitz1979}. 
To increase the sensitivity and to use a consistent approach while searching for a new generation of quarks, we have developed a simultaneous search for the up-type and down-type fourth-generation quarks,
based on both the electroweak and strong production mechanisms.

If a fourth generation of quarks exists, their production cross sections and decay branching fractions will be governed by an
extended $4\times 4$ CKM matrix, $V^{4\times 4}_{\mathrm{CKM}}$, in which we denote the up- and down-type fourth-generation quarks as \PQtpr and \PQbpr respectively.  For simplicity, we assume a model with one free parameter, $A$, where $0 \leq A \leq 1$:
\begin{align*}\label{eq:CKM}
        V_{\mathrm{CKM}}^{4\times4}
        &= \begin{pmatrix}
	  V_{\cPqu\cPqd}& V_{\cPqu\cPqs}  & V_{\cPqu\cPqb} & V_{\cPqu\PQbpr}\\
          V_{\cPqc\cPqd}& V_{\cPqc\cPqs}  & V_{\cPqc\cPqb} & V_{\cPqc\PQbpr}\\
          V_{\cPqt\cPqd}& V_{\cPqt\cPqs}  & V_{\cPqt\cPqb} & V_{\cPqt\PQbpr}\\
          V_{\PQtpr\cPqd}& V_{\PQtpr\cPqs}  & V_{\PQtpr\cPqb} & V_{\PQtpr\PQbpr}
        \end{pmatrix}\\
        &= \begin{pmatrix}
	  \mathcal{O}(1) & \mathcal{O}(0)  & \mathcal{O}(0) & 0 \\
          \mathcal{O}(0) & \mathcal{O}(1)  & \mathcal{O}(0) & 0 \\
          \mathcal{O}(0) & \mathcal{O}(0)  & \sqrt{A} & \sqrt{1-A} \\
          0 & 0  & -\sqrt{1-A} & \sqrt{A} \\
	\end{pmatrix}.
\end{align*}
The complex phases are not shown for clarity. Within this model, mixing is allowed only between the third and the fourth generations. This is a reasonable assumption since the mixing between the third and the first two generations is observed to be small~\cite{Nakamura}. However, the limits presented in this paper would be too stringent if there is a fourth generation that mixes only with the first two generations, or the size of the mixing with the third generation is about the same as the mixing with the first two generations.

With this search, we set limits on the masses of the fourth-generation quarks as a function of $A$. Since $\sqrt{A}=|V_{\cPqt\cPqb}|$, the lower limit of $|V_{\cPqt\cPqb}| > 0.81$ from the single-top production cross section measurements~\cite{Abazov:2012vd} translates into a lower limit on the mixing between the third- and fourth-generation quarks in our model of $A > 0.66$. 

Using the data collected from $\sqrt{s} = 7$\TeV proton-proton collisions at the Large Hadron Collider (LHC), we search for fourth-generation quarks that are produced in
pairs, namely \PQbpr{}\PAQbpr and \PQtpr{}\PAQtpr, or through electroweak production, in particular \cPqt{}\PQbpr, \PQtpr{}\cPqb, and \PQtpr{}\PQbpr, where the charges are omitted in the notation.
While the cross sections of the pair production processes do not depend on the value of $A$, the production cross sections of the \cPqt{}\PQbpr and \PQtpr{}\cPqb\ processes depend
linearly on $(1-A)$, and the single-top and \PQtpr{}\PQbpr cross sections on $A$.

We assume the \PQtpr and \PQbpr masses to be degenerate within 25\GeV. In the case they are degenerate, they will decay in 100\% of the cases to the third-generation quarks, since the decay of one fourth-generation quark to the other is kinematically not allowed.
However, even for non-zero mass differences, the branching
fractions of the $\PQtpr \rightarrow \cPqb\PW$ and the $\PQbpr \rightarrow \cPqt\PW \rightarrow (\cPqb\PW)\PW$ decays are close to 100\%, provided that the mass difference
is small~\cite{Chao:2011th}. For instance, for a mass splitting of 25\GeV, and for $V_{\PQtpr\PQbpr}=0.005$ (which would correspond to $A=0.99975$ in our model), less than 5\% of
the decays will be $\PQbpr \rightarrow \PQtpr\PW^{*}$ (in the case $m_{\PQtpr}<m_{\PQbpr}$) or  $\PQtpr \rightarrow \PQbpr\PW^{*}$ (in the case $m_{\PQtpr}>m_{\PQbpr}$). For larger values of $V_{\PQtpr\cPqb}$, the branching fractions of $\PQbpr \rightarrow \PQtpr\PW^{*}$ (or $\PQtpr \rightarrow \PQbpr\PW^{*}$) decrease even further. Therefore, the decay chains remain unchanged as long as the mass splitting is relatively small. 
We expect the following final states:
\begin{itemize}
\item $\PQtpr\cPqb \rightarrow \cPqb\PW\cPqb$;
\item $\PQtpr\PAQtpr \rightarrow \cPqb\PW\cPqb\PW$;
\item $\PQbpr\cPqt \rightarrow \cPqt\PW\cPqb\PW \rightarrow \cPqb\PW\PW\cPqb\PW$;
\item $\PQbpr\PQtpr \rightarrow \cPqt\PW\cPqb\PW \rightarrow \cPqb\PW\PW\cPqb\PW$;
\item $\PQbpr\PAQbpr \rightarrow \cPqt\PW\cPqt\PW \rightarrow \cPqb\PW\PW\cPqb\PW\PW$.
\end{itemize}
These decay chains imply that two jets from \cPqb\ quarks and one to four \PW\ bosons are expected in the final state for fourth-generation quarks produced both singly and in pairs.
The \PW\ bosons decay to either hadronic or leptonic final states. Events with either one isolated lepton (muon or electron) or two same-sign dileptons or three leptons are selected.
The different production processes are classified according to the number of observed \PW\ bosons.

\section{The Compact Muon Solenoid detector}
\label{sect:cms}
The central feature of the Compact Muon Solenoid (CMS) detector is a superconducting solenoid, 13\unit{m} in length and 6\unit{m} in internal diameter, providing an axial magnetic field of 3.8\unit{T}. The inside
of the solenoid is equipped with various particle detection systems. Charged particle trajectories are measured by a silicon pixel and strip tracker, covering
$0<\phi<2\pi$ in azimuth and $|\eta|<2.5$, where the pseudorapidity $\eta$ is defined as $-\ln[\tan(\theta/2)]$, and $\theta$ is the polar angle of the trajectory with
respect to the anticlockwise-beam direction. A crystal electromagnetic calorimeter (ECAL) and a brass/scintillator hadron calorimeter surround the tracking volume and
provide high-resolution energy and direction measurements of electrons, photons, and hadronic jets. Muons are measured in gas-ionization detectors embedded in the steel return yoke outside the solenoid. The CMS
detector also has extensive forward calorimetry covering up to $|\eta|<5$. The detector is nearly hermetic, allowing for energy balance measurements in the plane transverse to the beam directions. A two-tier trigger system selects the most interesting proton collision events for use in physics analysis. A more detailed description of the CMS detector can be found elsewhere~\cite{CMS:2008zzk}.
\section{Event selection and simulation}
\label{sect:simulation}
The search for the fourth-generation quarks is performed using the $\sqrt{s} = 7$\TeV proton-proton collisions recorded by the CMS experiment at the LHC.
We have analyzed the full dataset collected in 2011 corresponding to an integrated luminosity of $(5.0 \pm 0.1)$\fbinv. Events are selected with a
trigger requiring an isolated muon or electron, where the latter is accompanied by at least one jet identified as a \cPqb\ jet. The muon system, the calorimetry and the tracker are
used for the particle-flow (PF) event reconstruction~\cite{CMS-PAS-PFT-10-002}. Jets are reconstructed using the anti-\kt algorithm~\cite{cacciari-2008} with a size parameter of 0.5.
Events are further selected with at least one high-quality isolated muon or electron with a transverse momentum (\PT) exceeding 40\GeV in the acceptance
range $|\eta| < 2.1$ for muons and $|\eta| < 2.5$ for electrons. The relative isolation, $I_{\text{rel}}$, is calculated from the other PF particles
within a cone of $\Delta R = \sqrt{(\Delta\phi)^2+(\Delta\eta)^2}<0.4$ around the axis of the lepton. It is defined as $I_{\text{rel}} =
(E_{\mathrm{T}}^{\text{charged}}+E_{\mathrm{T}}^{\text{photon}}+E_{\mathrm{T}}^{\text{neutral}})/\PT$, where $E_{\mathrm{T}}^{\text{charged}}$ and $E_{\mathrm{T}}^{\text{photon}}$ are the transverse energies
deposited by charged hadrons and photons, respectively, and $E_{\mathrm{T}}^{\text{neutral}}$ is the transverse energy deposited by neutral particles
other than photons. We identify muons and electrons as isolated when $I_{\text{rel}}<0.125$ and $I_{\text{rel}}<0.1$, respectively. The requirement
on the relative isolation for electrons is tighter than for muons because the backgrounds for electrons are higher than for muons.
Electron candidates in the transition region between ECAL barrel and endcap ($1.44 < |\eta| < 1.57$) are excluded because
the reconstruction of an electron object in this region is not optimal. We require a missing transverse momentum
\ETslash of at least 40\GeV. The \ETslash is calculated as the absolute value of the vector sum of the \PT of
all reconstructed objects. Jets are required to have a $\PT > 30$\GeV. The jet energies are corrected to establish a
uniform response of the calorimeter in $\eta$ and a calibrated absolute response in \PT. Furthermore, a correction is applied to take into account the
energy clustered in jets due to additional proton interactions in the same bunch crossing.

The observed data are compared to simulated data generated with \textsc{powheg 301}~\cite{Alioli:2009je,Re:2010bp} for the single-top process,
\textsc{pythia 6.4.22}~\cite{Sjostrand:2006za} for the diboson processes, and \textsc{MadGraph 5.1.1}~\cite{Alwall:2011uj} for the signal and other standard model
processes. The \textsc{powheg} and \textsc{MadGraph} generators are interfaced with \textsc{pythia} for the decay of the particles as well as the hadronization and
the implementation of a CMS custom underlying event tuning (tune Z2)~\cite{Field:2010bc}. The matching of the matrix-element partons to the parton showers is
obtained using the MLM matching algorithm~\cite{Mangano:2006rw}. The \textsc{cteq6L1} leading-order (LO) parton distributions are used in the event
generation~\cite{Nadolsky:2008zw}. The generated events are passed through the CMS detector simulation based on \GEANTfour~\cite{Allison:2006ve}, and then processed by the same reconstruction software as the collision data. The simulated events are reweighted to match the observed distribution of the number of simultaneous proton interactions. For the full dataset collected in 2011, we observe on average about nine interactions in each event.    
We smear the jet energies in the simulation to match the resolutions measured with data~\cite{CMS-PAPER-JME-10-011}. At least one of the jets within the tracker acceptance ($|\eta|<2.4$) needs to be identified
as a \cPqb\ jet. For the \cPqb-jet identification, we require the signed impact parameter significance of the third track in the jet (sorted by decreasing significance) to be larger than a value chosen such that the probability for a light quark jet to be misidentified as a \cPqb\ jet is about 1\%.
We apply scale factors measured from data to the simulated events to take into account the different \cPqb-jet efficiency and the different probability that a light quark or gluon is identified as a \cPqb\ jet in data and simulation~\cite{BTV-11-004}.

The top-quark pair as well as the \PW\ and \Z production cross section values used in the analysis correspond to the measured values from CMS~\cite{PAPER-TOP-10-003,CMS:2011aa}.
We use the predicted cross section values for the single-top, $\ttbar+\PW$, $\ttbar+\Z$, and same-sign $\PW\PW$
processes~\cite{Kidonakis:2011wy,Kidonakis:2010tc,Kidonakis:2010ux,Hirschi:2011pa}. The cross section values for the diboson production are obtained with the MCFM generator~\cite{MCFM,vectorbosonpair}. 

For the pair-production of the fourth-generation quarks we use the approximate next-to-next-to-leading-order cross section values from~\cite{Hathor}.
For the electroweak production processes mentioned above, we rescale the next-to-leading-order (NLO) cross sections at
14\TeV~\cite{Campbell:2009gj} to 7\TeV using a scale factor defined as the ratio of the LO cross section at
7\TeV and the LO cross section at 14\TeV as obtained by the \textsc{MadGraph} event generator.
The resulting production cross sections are maximal, hence assuming $|V_{\cPqt\PQbpr}| =
|V_{\PQtpr\cPqb}| = |V_{\PQtpr\PQbpr}| = 1$, and are rescaled according to the value of $A$.

\section{Event classification}
\label{sect:classification}
Different channels are defined according to the number of \PW\ bosons in the final state. Given that the \PQtpr decay mode is the same as the top-quark decay
mode, the \PQtpr{}\cPqb\ and \PQtpr{}\PAQtpr processes will yield signatures that are very similar to respectively the single-top and \ttbar processes in the standard
model. We select these processes through the single-lepton decay channel.
In the signal final states that contain a \PQbpr quark, we expect three or four \PW\ bosons. If two or more of these \PW\
bosons decay to leptons, we may have events with two leptons of the same charge or with three charged leptons. Although the branching fraction of these decays is small compared to that of other decay channels, these final states are very interesting because of the low background that is expected from standard model processes.

\subsection{The single-electron and single-muon decay channels}
On top of the aforementioned event selection criteria, we veto events with additional electrons or muons with $I_{\mathrm{rel}}< 0.2$ and $\PT>10$\GeV for muons and
$\PT>15$\GeV for electrons. We divide the selected single-lepton events into different subsamples according to the signal final states. Therefore,
we define a procedure to count the number of \PW-boson candidates. Each event has at least one \PW\ boson that decays to leptons, consistent with the
requirements of an isolated lepton and a large missing transverse momentum from the neutrino that escapes detection. The decays of \PW\ bosons to,
$\cPq\cPaq$ final states are reconstructed with the following procedure. For each event, we have a collection of selected jets used as input for the
reconstruction of the \PW-boson candidates. The one or two jets that are identified as \cPqb\ jets are removed from the collection. \PW-boson candidates are constructed from all possible pairs of the remaining jets in the collection. We use both the expected mass, $m^\text{fit}_\PW
= 84.3$\GeV, and the width, $\sigma^\text{fit}_{m_\PW} = 9.6$\GeV from a Gaussian fit to the reconstructed mass distribution of jet pairs from the decay of a \PW\ boson in
simulated \ttbar events. The \PW-boson candidate with a mass that matches the value of $m^\text{fit}_\PW$ best, is chosen as a \PW\ boson if its mass is
within a $\pm 1 \sigma^\text{fit}_{m_\PW}$ window around $m^\text{fit}_\PW$. The jet pair that provided the hadronically decaying \PW\ boson is removed from the collection and the
procedure is repeated until no more candidates are found for \PW\ bosons decaying to jets. 
Different exclusive subsamples are defined according to the number of \cPqb\ jets (exactly one or at least two) and the number of \PW-boson candidates (one,
two, three, and at least four). There are seven subsamples, because we do not consider the subsample with only one \cPqb\ jet and one \PW\ boson. The
subsample with two \cPqb\ jets and one \PW\ boson is dominated by singly produced \PQtpr events. In this subsample, we apply a veto for additional jets with a
transverse momentum exceeding 30\GeV. Furthermore, since \bbbar background tends to have jets that are produced back-to-back with
balanced \PT, we remove this background by requiring $\Delta\phi(j_{1},j_{2})<\frac{\pi}{2}+\pi(|\PT^{j1}-\PT^{j2}|)/(\PT^{j1}+\PT^{j2})$.

Table~\ref{tab:overview1} summarizes the requirements that define the different single-lepton decay subsamples, after the criteria on the \ETslash, and the lepton and jet \PT and $\eta$ are applied.
\begin{table*}[hbtp]
\begin{center}
\topcaption{Overview of the event selection requirements defining the different subsamples in the single-lepton decay channel. The single-lepton decay channel is divided in seven different subsamples according to the number of \cPqb\ jets and the number of \PW-boson candidates.}
\label{tab:overview1}
\begin{tabular}{c|c|c|c}
\hline
\hline
\multicolumn{4}{c}{single-lepton decay channel} \\
1 \PW\  & 2 \PW\ & 3 \PW\ & 4 \PW\ \\
\hline
$=2$ jets & $\geq 4$ jets & $\geq 6$ jets & $\geq 8$ jets \\
$=2$ \cPqb\ jets & \multicolumn{3}{c}{either $= 1$ or $\geq 2$ \cPqb\ jets}  \\
$\Delta\phi(j_{1},j_{2})$ requirement \qquad & 1 $\PW \rightarrow \cPq\cPaq$ \qquad & 2 $\PW
\rightarrow \cPq\cPaq$ \qquad & 3 $\PW \rightarrow \cPq\cPaq$\\
\hline
\hline
\end{tabular}
\end{center}
\end{table*}

Table~\ref{tab:singlelepyields} shows the observed and predicted event yields. After the selection criteria, the dominant background contributions result from the production of top quark pairs, \PW+jets, and single top. Other processes with very small contributions to the total background are \Z+jets and diboson production, and also top quark pairs produced in association with a \PW\, or \Z boson. The combined event yield of these processes is about 1\% of the total standard model contribution. The multijet background is found to be negligible in each of the subsamples. The reason is the requirements of an isolated muon or electron with $\PT > 40$\GeV, a missing transverse momentum of 40\GeV and at least one jet identified as a b-jet. Data and simulation are found to agree within the combined statistic and systematic uncertainties. 

\begin{table*}[hbtp]
\begin{center}
\addtolength{\tabcolsep}{-2pt}
\caption{Event yields in the single lepton channel. Uncertainties reflect the combined statistical and systematic uncertainties. The prediction for the signal is shown for two different values of $A$ and for a fourth-generation-quark mass $m_{\cPq'}=550$\GeV.}
\label{tab:singlelepyields}
\begin{tabular}{cccccccc}
\hline
\hline
 & 1b 2W & 1b 3W & 1b 4W & 2b 1W & 2b 2W & 2b 3W & 2b 4W \\
\hline \\ [-2ex]
\ttbar+jets & $5630 \pm 410$  & $230^{+29}_{-26}$  & $3.0^{+1.9}_{-1.3}$  & $819^{+59}_{-62}$  & $2810 \pm 240$  & $85^{+12}_{-10}$  & $0.6^{+0.8}_{-0.5}$  \\ [0.5ex]
\PW+jets & $490 \pm 180$  & $8.0^{+3.1}_{-3.0}$ & $0.3^{+0.9}_{-0.3}$  & $150^{+47}_{-46}$ & $37 \pm 12$  & $1.1^{+1.0}_{-0.4}$  & $0.0^{+0.8}_{-0.0}$  \\ [0.5ex]
\Z+jets & $36^{+5}_{-6}$  & $1.0^{+0.2}_{-0.1}$  & $0$  & $7.1^{+1.0}_{-0.6}$  & $2.8^{+1.0}_{-0.3}$  & $0$  & $0$  \\ [0.5ex]
single top & $346 \pm 64$  & $6.5^{+1.6}_{-1.5}$  & $0.2^{+0.3}_{-0.2}$  & $200 \pm 34$  & $110 \pm 19$  & $2.5^{+0.7}_{-0.5}$  & $0.0^{+0.1}_{-0.0}$  \\ [0.5ex]
VV & $15 \pm 2$  & $0.4^{+0.3}_{-0.1}$ & $0.0^{+0.1}_{-0.0}$  & $15 \pm 2$  & $1.8 \pm 0.3$  & $0.0^{+0.1}_{-0.0}$  & $0.0^{+0.1}_{-0.0}$  \\ [0.5ex]
\ttbar V & $28 \pm 3$  & $3.4 \pm 0.5$  & $0.1 \pm 0.0$  & $0.7 \pm 0.2$  & $15 \pm 5$  & $1.5^{+0.3}_{-0.2}$  & $0$  \\ [0.5ex]
\hline \\ [-2ex]
Total background & $6550 \pm 450$  & $249^{+29}_{-26}$  & $3.6^{+2.1}_{-1.3}$  & $1190^{+83}_{-85}$  & $2970 \pm 240$  & $91^{+12}_{-10}$  & $0.6^{+1.2}_{-0.5}$  \\ [0.5ex]
Observed & $7003$  & $242$  & $8$  & $1357$  & $3043$  & $91$  & $4$  \\ [0.5ex]
Signal ($A=1$) & $55 \pm 1 $ & $12 \pm 1$  & $0.9 \pm 0.2$ & $1.0^{+0.2}_{-0.3}$  & $49 \pm 2$  & $8.1 \pm 0.4$  & $0.5 \pm 0.2$  \\ [0.5ex]
Signal ($A=0.8$) & $85 \pm2 $ & $14 \pm 1$  & $1.0 \pm 0.2$ & $69 \pm 3$  & $66 \pm 2$ & $9.2 \pm 0.4$  & $0.5 \pm 0.2$ \\ 
\hline
\hline
\end{tabular}
\end{center}
\end{table*}

\subsection{The same-sign dilepton and trilepton decay channels}
The transverse momentum of at least one of the leptons in the multilepton channel is required to be larger than 40\GeV, while the threshold is reduced to
20\GeV for additional leptons. Events with two muons or electrons with a mass within 10\GeV of the \Z-boson mass are rejected to reduce the standard
model background with \Z bosons in the final state. We require at least four jets for the same-sign dilepton events. In the case of the trilepton events
the minimum number of required jets is reduced to two. Table~\ref{tab:overview2} summarizes the event selection requirements defining the same-sign dilepton and trilepton decay
channels that are applied on top of the other requirements on the \ETslash, and lepton and jet \PT and $\eta$.
\begin{table*}[hbtp]
\begin{center}
\topcaption{Overview of the event selection requirements specific to the same-sign dilepton and trilepton decay channels.}
\label{tab:overview2}
\begin{tabular}{cc}
\hline
\hline
same-sign dilepton & trilepton \\
\hline
$=2$ isolated leptons with same sign & $=3$ isolated leptons \\
$\geq 4$ jets ($\PT> 30\GeV$, $|\eta|<2.4$) & $\geq 2$ jets ($\PT> 30\GeV$, $|\eta|<2.4$) \\
$\geq$ 1 \cPqb\ jet &  $\geq$ 1 \cPqb\ jet \\
\hline
\hline
\end{tabular}
\end{center}
\end{table*}

There are several contributions to the total standard model background for the same-sign dilepton
events. One of these contributions comes from events for which the charge of one of the leptons is misreconstructed, for instance in \ttbar events with two \PW bosons decaying into leptons. Secondly, there are events with one prompt
lepton and one non-prompt lepton passing the isolation and identification criteria. Finally, there is an irreducible contribution from standard model
processes with two prompt leptons of the same sign; \eg $\PW^{\pm}\PW^{\pm}$, $\PW\Z$, $\Z\Z$, $\ttbar + \PW$ and $\ttbar + \Z$.
Except for $\PW^{\pm}\PW^{\pm}$, these processes are also the main contributions to the total background for the trilepton subsample. The event yields for the irreducible component of the background for the same-sign dilepton channel and the total background in the case of the trilepton subsample are taken from the simulation. We obtain from the data the predicted number of background events for the first two contributions to the total background in the same-sign dilepton subsample.

For the same-sign dilepton events with at least one electron, the background is estimated from control samples. 
We determine the charge misidentification rate for electrons using a double-isolated-electron trigger. We require two isolated electrons with the
dielectron invariant mass within 10\GeV of the \Z-boson mass. We select events with $\ETslash < 20$\GeV and a transverse mass $M_\mathrm{T} = \sqrt{2
p_{\mathrm{T}}^{\ell}\ETslash}[1-\cos(\Delta\phi(\ell, \ETslash))]$ 
less than 25\GeV to suppress background from top-quark and \PW+jets events. We define the charge misidentification ratio $R$ as the number of events with two electrons of the same sign divided by twice the number of events with two electrons of opposite sign, i.e. $R=N_{SS}/2N_{OS}$. We obtain 0.14\% and 1.4\% for barrel and endcap electron candidates, respectively. After the full event selection is applied, with the exception of the electron sign requirement, we obtain a number of selected data events with two electrons and with an electron and a muon in the final state. The background with two electrons or with an electron and a muon with the same sign is obtained by taking the number of opposite-sign events and scaling it with $R$. The \PT spectrum of the electrons in the control sample and the signal region is similar. Therefore, no correction is applied for the \PT dependency of the charge misidentification ratio.

Another important background contribution to the same-sign dilepton channel originates from jets being misidentified as an electron or a muon (``fake'' leptons). Two collections of leptons, ``loose'' and ``tight'', are defined based on the isolation and identification criteria. Loose leptons are required to fulfill $I_{\text{rel}}< 0.2$, in contrast with $I_{\text{rel}}< 0.125 (0.1)$ for tight muons (electrons). Moreover, we require $|\eta|<2.5$ and $\PT<10 (15)$ for loose muons (electrons). Additionally, several identification criteria, that are intended to ensure the compatibility of the lepton track with the primary vertex, are relaxed. We require at least one loose electron or muon. Additionally, we require $\ETslash < 20$\GeV and
$M_{T} < 25$\GeV to suppress background from top-quark and \PW+jets events. Moreover, we veto events with leptons of the same flavor that have a dilepton mass within 20\GeV of the \Z boson mass. We count the number of loose and tight leptons with a \PT below 35\GeV. The threshold on the \PT is required to suppress contamination from \PW+jets events that would bias the estimation, because leptons produced in jets have typically a soft \PT spectrum. The probability that a loose (L) lepton passes the tight (T) selection criteria is then given by the ratio $\epsilon_{TL} = N_{T}/N_{L}$. To estimate the number of events from the background source with a non-prompt lepton, we count the number of events in data that pass the event selection criteria with one lepton passing the tight selection criteria and a second lepton passing the loose, but not the tight, criteria. This yield is multiplied by $\epsilon_\mathrm{TL}(1-\epsilon_\mathrm{TL})$ to determine the number of events with a non-prompt lepton in the analysis. The statistical uncertainty on the estimated number of events is large because only a few events are selected with one tight and one loose, but not tight, lepton.

The total number of expected background events for the same-sign dilepton and trilepton channels is given in Table~\ref{tab:total}.
\begin{table*}[hbtp]
\begin{center}
\topcaption{The prediction for the total number of background events compared with the number of observed events in the same-sign dilepton and the trilepton subsamples. The numbers of expected signal events are also shown for two possible scenarios.}
\label{tab:total}
\begin{tabular}{lcccc}
\hline
\hline
type  & 2 muons & 2 electrons & electron+muon & trilepton \\
\hline
Irreducible background &	$0.77 \pm 0.08$  &	$0.59 \pm 0.08$ &	$1.10 \pm 0.11$ &	  $0.96 \pm 0.12$  \\
Background from charge misid   & 	$-$	&	$0.47 \pm 0.08$ &	$0.71 \pm 0.06$ &	$-$    \\
Background from fake leptons   &	$0.06 \pm 0.06$ &	$0.30 \pm 0.15$ & $0.46 \pm 0.17$	 &  $-$  \\
\hline
Total background  & $0.83 \pm 0.11$ & $1.36 \pm 0.19$& $2.27 \pm 0.22$ & $0.96 \pm 0.12$  \\
Observed & 2  & 2  & 2  & 1 \\
Signal ($A=1$, $m_{\cPq'}=550$\GeV) & $3.31 \pm 0.15$ & $2.03 \pm 0.36$	& $5.29 \pm 0.19$	& $3.37 \pm 0.16$\\	
Signal ($A=0.8$, $m_{\cPq'}=550$\GeV) & $3.79 \pm 0.15$ & $2.29 \pm 0.36 $	& $6.00 \pm 0.19$	& $3.65 \pm 0.16$\\
\hline
\hline
\end{tabular}
\end{center}
\end{table*}

\section{Setting lower limits on the fourth-generation quark masses} 
\label{sect:lowerlimits}
We have defined different subsamples according to the reconstructed final state. In each of the different subsamples, we reconstruct observables that are
sensitive to the presence of the fourth-generation quarks. These observables are used as input to a fit of the combined distributions for the standard model (background-only) hypothesis and the signal-plus-background hypothesis. With the profile likelihood ratio as a test statistic, we calculate the 95\% confidence level (CL) upper limits on the combined input cross section of the signal as a function of the $V_{\mathrm{CKM}}^{4\times4}$ parameter $A$ and the mass of the fourth-generation quarks.

\subsection{Observables sensitive to the fourth-generation quark production}

The expected number of events is small in the subsamples with two leptons of the same sign, the trilepton subsample, and the two single-lepton subsamples with four \PW-boson candidates. As a consequence, the event counts in each of these subsamples are used as the observable. Table~\ref{tab:total} summarizes the
event counts for the subsamples with two leptons of the same sign and the trilepton subsample.

\begin{figure*}[hbtp] 
  \centering
  \includegraphics[width=0.49 \textwidth]{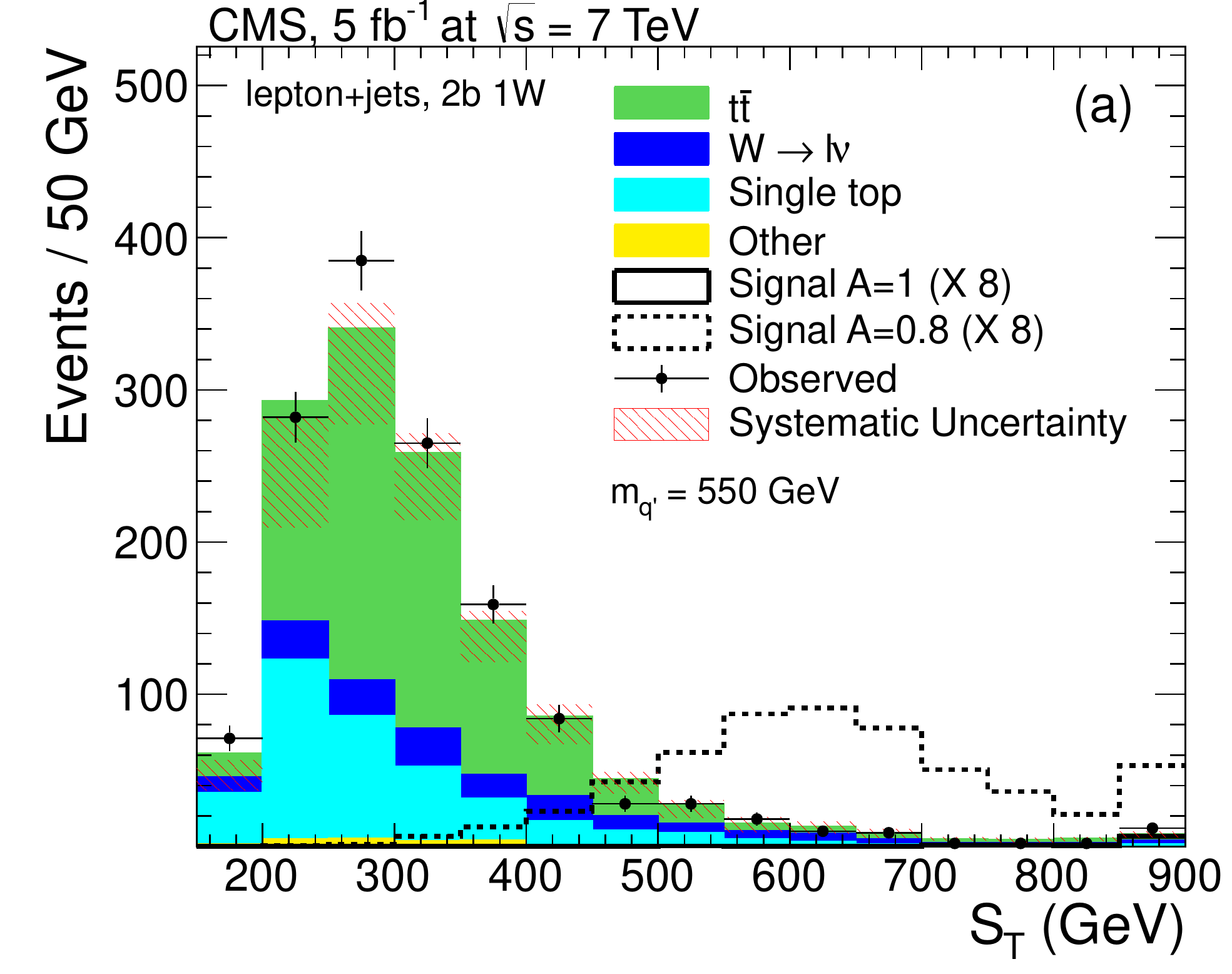}
  \includegraphics[width=0.49 \textwidth]{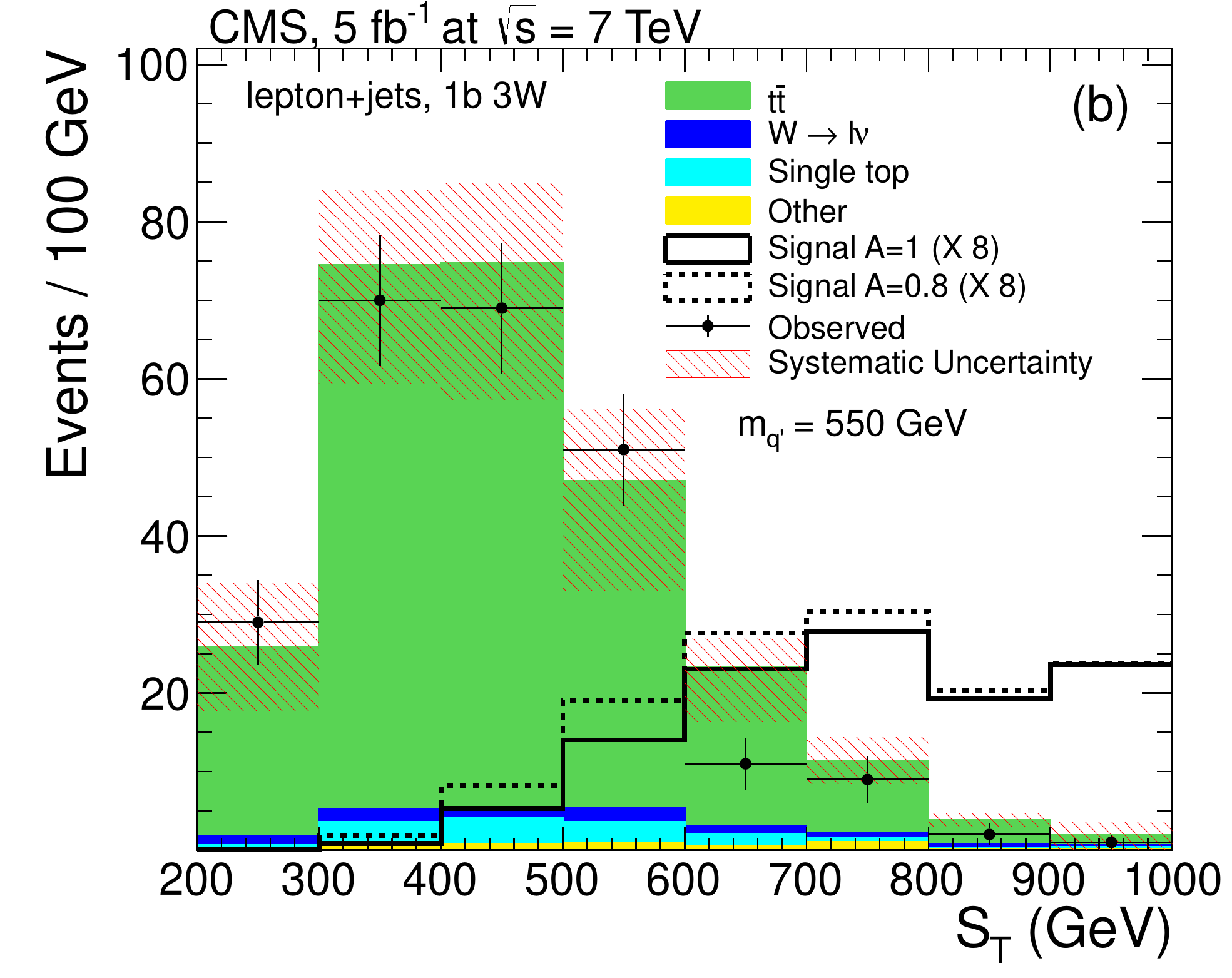}  \\
\vspace{0.3cm}
  \includegraphics[width=0.49 \textwidth]{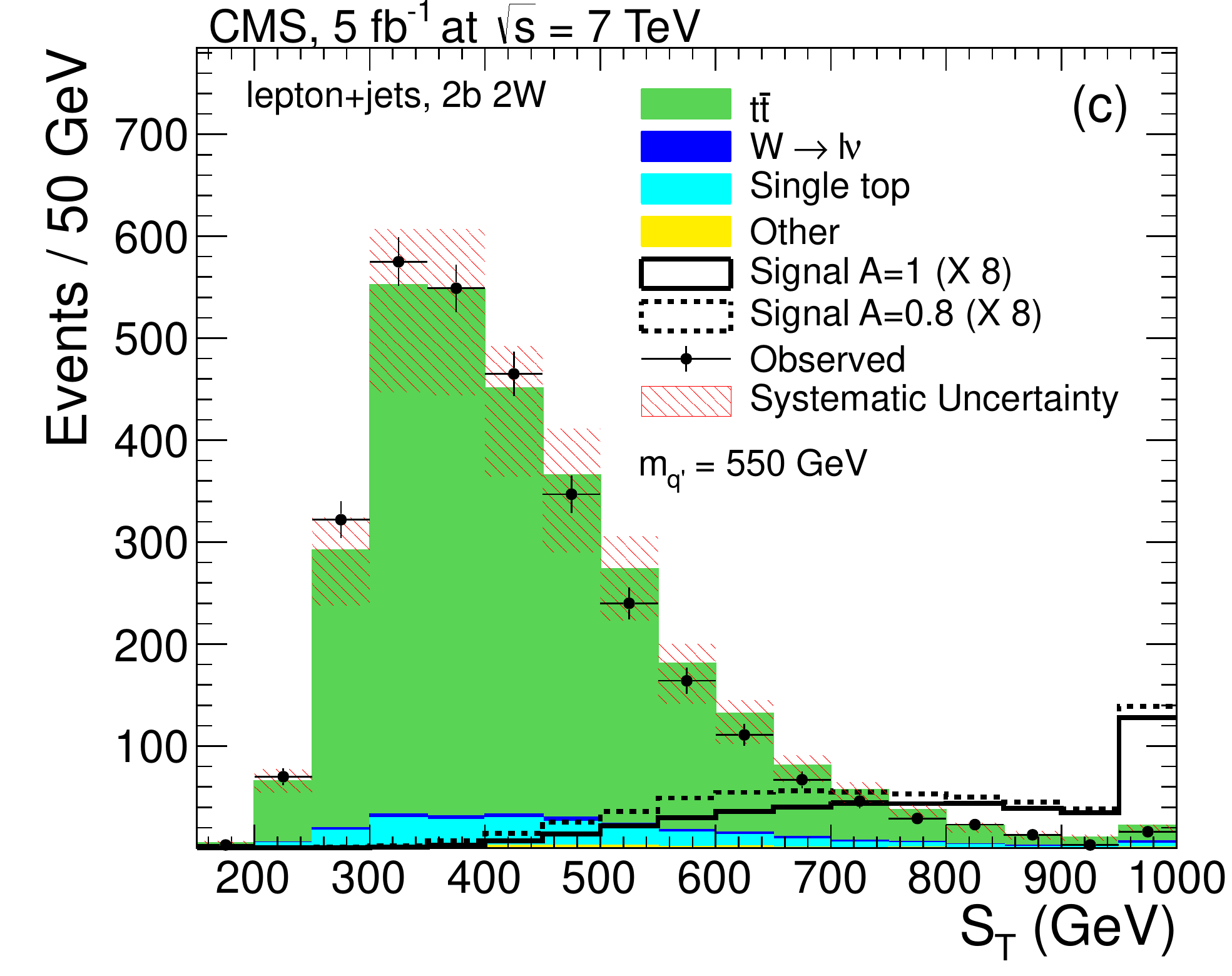}
  \includegraphics[width=0.49 \textwidth]{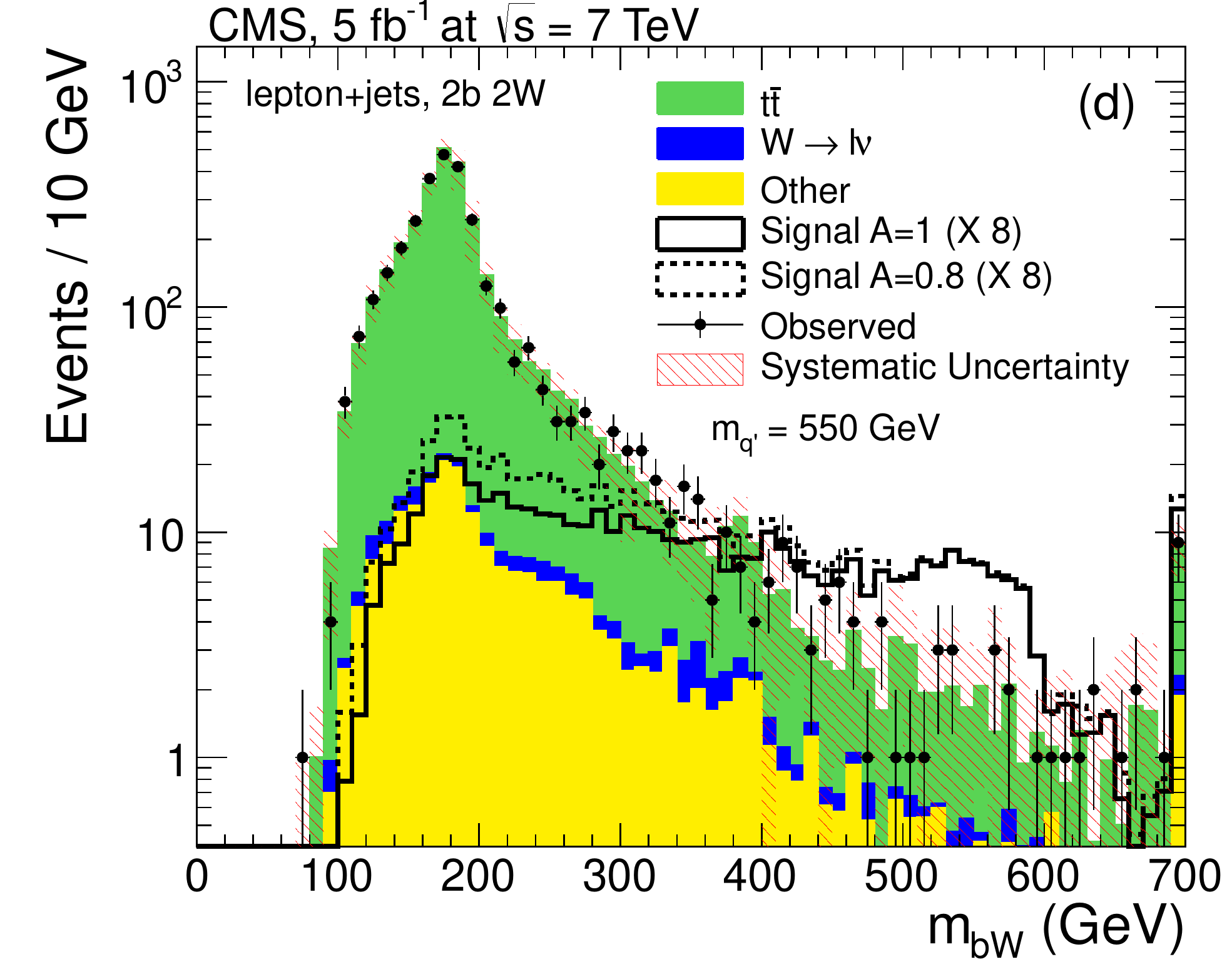}
  \caption{The $S_{\mathrm{T}}$ distribution for the subsamples with two \cPqb\ jets and one \PW\ boson (a), one \cPqb\ jet and three \PW\ bosons
  (b), two \cPqb\ jets and two \PW\ bosons (c), and the $m_{\cPqb\PW}$ distribution for the subsample with two \cPqb\ jets and two \PW\
  bosons (d). The data distributions of these observables are compared to their expectation from the simulation assuming the fitted
  nuisance parameters. The fitted values of the nuisance parameters represent the systematic shifts that are applied on the simulation to fit the data in the background-only hypothesis.
  As an illustration the total uncertainty band is shown around the simulated expected distribution before taking into account the fitted values of the nuisance parameters. The expected distribution
  for a signal is shown for two different values of the $V_{\mathrm{CKM}}^{4\times4}$ parameter $A$ and for \PQbpr and \PQtpr masses of 550\GeV.
  The cross section of the signal in the plots is scaled by a factor of eight for visibility. The last bin in all the histograms includes the overflow.
  We do not expect much signal for $A = 1$ in (a), because the subsample with two \cPqb\ jets and one \PW\ boson is mainly sensitive to single \PQtpr quark production.
  \label{fig:ST}
  }
\end{figure*}

In the single-lepton subsamples with one or three \PW\ bosons, we use $S_{\mathrm{T}}$ as the observable to discriminate between the standard model
background and the fourth-generation signal, where $S_{\mathrm{T}}$ is defined as the scalar sum of the transverse momenta of the reconstructed objects in
the final state, namely:
\begin{equation}
S_{\mathrm{T}} = \ETslash + \PT^{\ell} + \PT^{b} + \PT^{j} + \sum_{i=0}^{N} \PT^{W_{q\bar{q}}^{i}},
\label{eq:ST}
\end{equation}
where the sum runs over the number of reconstructed hadronically decaying \PW\ bosons; $\PT^{\ell}$ is the \PT of the
lepton, $\PT^{b}$ the \PT of the \cPqb\ jet, $\PT^{j}$ the \PT of the second \cPqb\ jet or, if
there is no additional jet identified as a \cPqb\ jet, the \PT of the jet with the highest transverse momentum in the event that
is not used in the \PW-boson reconstruction, and $\PT^{W_{q\bar{q}}^{i}}$ the \PT of the
$i^{{\rm th}}$ reconstructed \PW\ boson decaying to jets. In general, the decay products of the fourth-generation quarks are
expected to have higher transverse momenta compared to the standard model background. This is shown in Fig.~\ref{fig:ST} for three of the
subsamples. The dominant contribution to the selected signal events in the subsample with two \cPqb\ jets and one \PW\ boson would come from the \PQtpr{}\cPqb\ process. Almost no signal events are selected for $A=1$, because in that case the production cross section of \PQtpr{}\cPqb\ is equal to zero.
The subsamples with two \PW\ bosons are dominated by \ttbar events. In this case we use two sensitive observables, $S_{\mathrm{T}}$ 
 and the mass of the hadronic \cPqb{}\PW\ system, $m_{\cPqb\PW}$. The latter observable is sensitive to the fourth generation physics, because of the higher mass of a
 hypothetical fourth generation \PQtpr quark compared to the top-quark mass. To obtain a higher sensitivity with the $m_{\cPqb\PW}$ observable, four jets need to be
 assigned to the quarks to reconstruct the final state $\PQtpr\PAQtpr \rightarrow \PW\cPqb\PW\cPqb \rightarrow  \cPq \cPaq \cPqb \ell \nu_{\ell} \cPqb$. Therefore, six observables with
 discriminating power between correct and wrong jet/quark assignments are combined with a likelihood ratio method. These observables are angles
 between the decay products, the \PW-boson mass, the transverse momentum of the top quark decaying to hadrons, and an observable related to the values of the \cPqb-jet
 identification variable for the jets. The jet/quark assignment with the largest value of the likelihood ratio is chosen. The mass of the \cPqb{}\PW\ system is then reconstructed
 from this chosen jet/quark assignment. 
The lower plots in Fig.~\ref{fig:ST} show the projections of the two-dimensional $S_{\mathrm{T}}$ versus $m_{\cPqb\PW}$ distribution.

An overview of the observables used in the fit for the presence of the
fourth-generation quarks is presented in Table~\ref{tab:overviewobservables}.
\begin{table}[hbtp]
\begin{center}
\topcaption{Overview of the observables used in the limit calculation.}
\label{tab:overviewobservables}
\begin{tabular}{cc}
\hline
\hline
subsample & observable \\
\hline
single-lepton 1\PW &  $S_{\mathrm{T}}$ \\
single-lepton 2\PW &  $S_{\mathrm{T}}$ and $m_{\cPqb\PW}$ \\
single-lepton 3\PW &  $S_{\mathrm{T}}$ \\
single-lepton 4\PW &  event yield \\
same-sign dilepton &  event yield \\
trilepton &  event yield \\
\hline
\hline
\end{tabular}
\end{center}
\end{table}

\subsection{Fitting for the presence of fourth-generation quarks}

We construct a single histogram ``template'' that contains the information of the sensitive observables from all the subsamples. Different
template distributions are made for the signal corresponding to the different values of $A$ and the fourth-generation quark masses $m_{\cPq'}$. The
binning of the two-dimensional observable distribution in the single-lepton subsamples with two \PW\ bosons is defined using the following
procedure. We use a binning in the dimension of $m_{\cPqb\PW}$ such that the top-quark pair background events are uniformly distributed
over the bins. Secondly, the binning in the dimension of $S_{\mathrm{T}}$ in each of the $m_{\cPqb\PW}$ bins is chosen to obtain
uniformly distributed top-quark pair events also in this dimension. 

The templates of the sensitive observables are used as input to obtain the likelihoods for the background-only and the signal-plus-background hypotheses.
Systematic uncertainties are taken into account by introducing nuisance parameters that may affect the shape and the normalization of the
templates. In a case where the systematic uncertainty alters the shape of the templates, template morphing~\cite{Read:1999kh,Conway-PhyStat} is used to interpolate linearly on a bin-by-bin basis between the nominal templates and systematically shifted ones.

The normalization of the templates is affected by the uncertainty in the integrated luminosity, the lepton efficiency and the normalization of the background processes.
The integrated luminosity is measured with a precision of 2.2\%~\cite{Lumi} and has the same normalization effect on all the templates.
The uncertainties in the lepton efficiency are a combination of the trigger, selection and identification efficiencies that amount to 3\% and 5\% for muon and electron respectively.
For the uncertainty in the normalization of the background processes, we use the uncertainties in the production cross section of the various standard model processes.
The most important contributions that affect the normalization of the templates are the 12\%~\cite{PAPER-TOP-10-003} (30\%) uncertainty for the top-quark pair (single-top)
production cross section and a 50\% uncertainty for the \PW\ production cross section because of the large fraction of selected events with jets from heavy flavor
quarks. For the multilepton channel, we take into account the uncertainties in the background estimation obtained from the data.
We also include the uncertainties in the production cross sections of \Z (5\%~\cite{CMS:2011aa}), \PW{}\PW\ (35\%), \PW{}\Z (42\%), \Z{}\Z (27\%), $\ttbar+W$
(19\%), $\ttbar+Z$ (28\%) and $\PW^{\pm}\PW^{\pm}$ (49\%). The uncertainties in the normalization of diboson and top quark pair production in association with a boson are
taken from a comparison of the NLO and the LO predictions.

The largest systematic effects on the shape of the templates originate from the jet energy corrections~\cite{CMS-PAPER-JME-10-011} and the scale factors between
data and simulation for the \cPqb-jet efficiency and the probability that a light quark or gluon is identified as a \cPqb\ jet~\cite{BTV-11-004}. These effects are
estimated by varying the nominal value by ${\pm}1$ standard deviation. The uncertainty in the jet energy resolution of about 10\% has a relatively small effect on the
expected limits. The same is true for the uncertainty in the modeling of multiple interactions in the same beam crossing. The latter effect is evaluated by varying the
average number of interactions in the simulation by 8\%.

The probability density functions of the background-only and the signal-plus-background hypothesis are fitted to the data to fix the nuisance parameters in both models. In the signal-plus-background model, an additional variable, defined as the cross section for the fourth-generation signal obtained by combining the separate search channels, is included. In the combined cross section variable the relative fraction of each fourth-generation signal process is fixed according to the
probed model parameters $(A,m_{\cPq'})$. 
Using a Gaussian approximation for the probability density function of the test statistic, we determine the 95\% CL expected and observed limits on the combined cross section variable using the $CL_{s}$ criterion~\cite{asymptoticCLs,Junk1999435,0954-3899-28-10-313}.
We exclude the point $(A,m_{\cPq'})$ at the 95\% CL if the upper limit on the combined cross section variable is smaller than its
predicted value within the fourth-generation model. The procedure is repeated for each value of $A$ and $m_{\cPq'}$.

\subsection{Results and discussion}
\begin{figure}[hbtp]
\centering
   \includegraphics[width=0.45\textwidth]{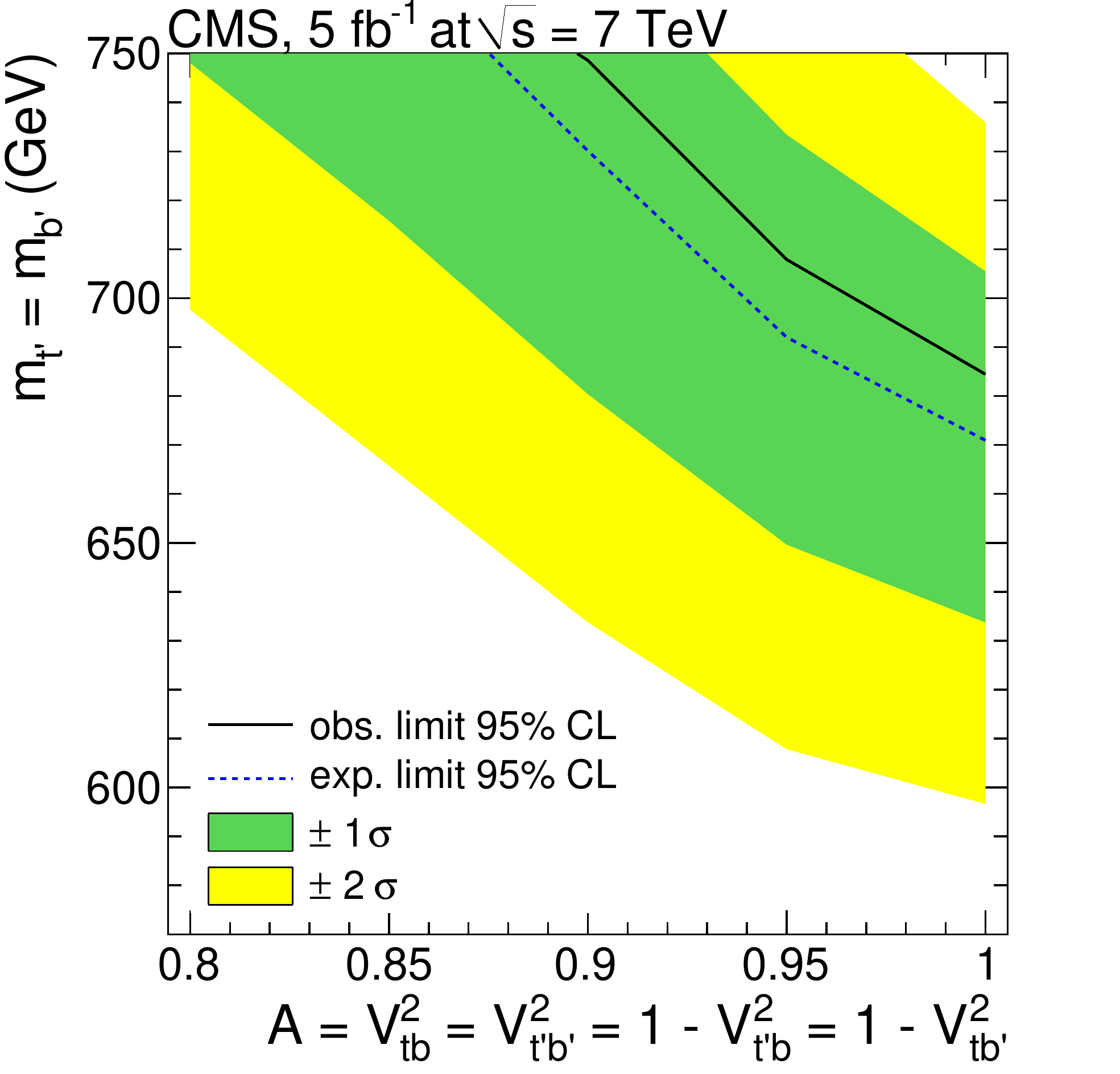}
   \label{fig:CLsLimit_Combined_massdegenerate}\\
\vspace{0.3cm}
   \includegraphics[width=0.45\textwidth]{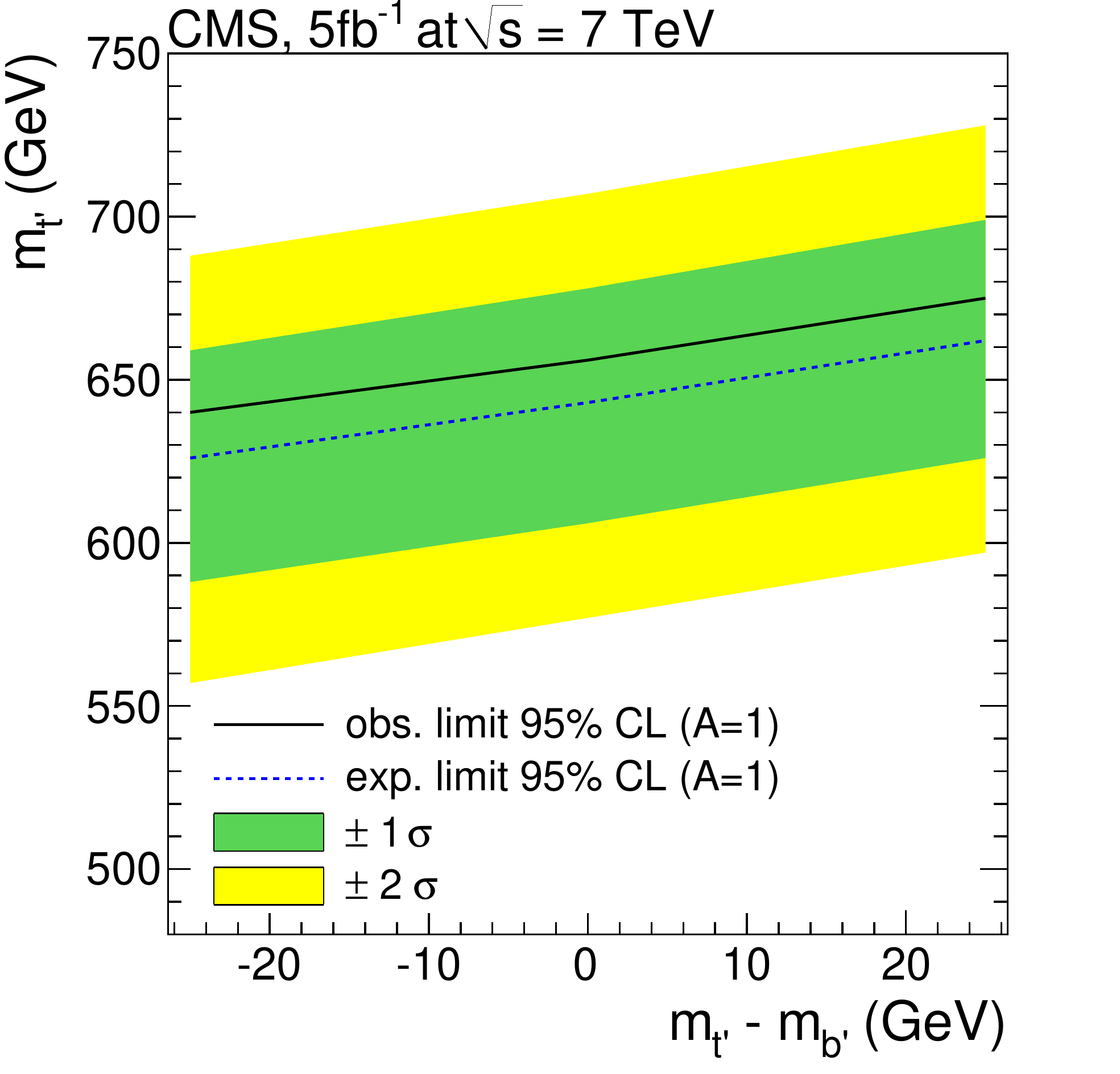}
   \label{fig:CLsLimit_Combined_notmassdegenerate}
   \caption{Top: Exclusion limit on $m_{\PQtpr} = m_{\PQbpr}$ as a function of the $V_{\mathrm{CKM}}^{4\times4}$ parameter $A$. The
   parameter values below the solid line are excluded at 95\% CL. The inner (outer) band indicates
   the 68\% (95\%) confidence interval around the expected limit. The slope indicates the sensitivity
   of the analysis to the \PQtpr{}\cPqb\ and \cPqt{}\PQbpr processes. Bottom: For
   a $V_{\mathrm{CKM}}^{4\times4}$ parameter value $A\sim1$, the exclusion limit on $m_{\PQtpr}$ versus $m_{\PQtpr}-m_{\PQbpr}$ is shown. The exclusion limit is calculated for
   mass differences up to 25\GeV. The existence of up-type fourth-generation quarks with mass values below the observed limit are excluded at the 95\% CL.}
\label{fig:CLsLimit_Combined}
\end{figure}

We use the $CL_{s}$ procedure to calculate the combined limit for the single muon, single electron, same-sign dilepton and
trilepton channels. When the value of the $V_{\mathrm{CKM}}^{4\times4}$ parameter $A$ approaches unity, the standard model single-top and the \PQtpr{}\PQbpr processes reach their maximal values for the production cross section. When the value of $A$ decreases, the cross
section of these processes decreases linearly with $A$. At the same time the expected cross section of the \PQtpr{}\cPqb\
and \cPqt{}\PQbpr processes increases with $(1-A)$ and is equal to zero for $A=1$. Therefore, the \PQtpr{}\cPqb\ and \cPqt{}\PQbpr processes are expected to enhance the sensitivity for fourth-generation quarks when the parameter $A$ decreases. This is visible in
the upper part of Fig.~\ref{fig:CLsLimit_Combined} where both the expected and observed limits on $m_{\cPq'}$ are more stringent for
smaller values of $A$. For instance, the limit on the fourth-generation quark masses increases by 70\GeV for $A=0.9$ compared to the value of the limit for $A\sim1$. While the \PQtpr{}\cPqb\ and \cPqt{}\PQbpr processes do not contribute for $A\sim1$, the inclusion of the \PQtpr{}\PQbpr process results in a more stringent limit (a difference of about 30\GeV) compared to when this process is not taken into account.

The existence of fourth-generation quarks with degenerate masses is excluded for all parameter values
below the line using the assumed model of the $V_{\mathrm{CKM}}^{4\times4}$ matrix. In particular, fourth-generation quarks with a degenerate mass below 685\GeV are excluded at the 95\% CL for a parameter value of $A\sim1$. It is worth noting that no limits can be set for $A$ exactly equal to unity ($A=1$), because in this special case the fourth-generation quarks would be stable in the assumed model.   
The analysis is however valid for values of A extremely close to unity. The distance between the primary vertex and the decay vertex of the fourth-generation quarks
is less than 1\mm for $1-A > 2\ten{-14}$, a number obtained using the LO formula for the decay width of the top quark in which the top-quark mass is replaced
with a fourth-generation-quark mass of 600\GeV.

Up to now, the masses of the fourth-generation quarks were assumed to be degenerate. However, if a fourth generation of chiral quarks exists, this is not necessarily the
case. Therefore, it is interesting to study how the limit would change for non-degenerate quark masses. If we assume non-degenerate masses, another decay channel for the
fourth-generation quarks is possible. Namely, the branching fraction for the decay of \PQtpr (\PQbpr)
into \PQbpr (\PQtpr) and an off-shell \PW\ boson becomes non-zero. For values of the mass splitting up to about 25\GeV, this branching
fraction is small as noted in the introduction. We assume a mass splitting of 25\GeV and unchanged branching fractions for the \PQtpr and \PQbpr decays. The sensitivity of the analysis increases or decreases depending on the specific values of the masses and hence the production cross sections of the fourth-generation quarks. 
The effect of the mass difference between the fourth-generation quarks on the exclusion limit is shown in the bottom plot of Fig.~\ref{fig:CLsLimit_Combined} for
a $V_{\mathrm{CKM}}^{4\times4}$ parameter $A\sim 1$. For instance in case $m_{\PQtpr}=m_{\PQbpr} + 25\GeV$ ($m_{\PQtpr}=m_{\PQbpr} - 25\GeV$), the limit on $m_{\PQtpr}$ increases about $+20$ ($-20$)\GeV
with respect to the degenerate-mass case. To obtain this limit, we do not take into account the electroweak \PQtpr{}\PQbpr process, which results in more conservative exclusion limits. In particular one observes that quarks with degenerate masses below about 655\GeV are excluded at the 95\% CL compared to 685\GeV when the \PQtpr{}\PQbpr process is included.

\section{Summary}
\label{sect:results}

Results from a search for a fourth generation of quarks have been presented. A simple model for a unitary CKM matrix has been defined based on a
single parameter $A = |V_{\cPqt\cPqb}|^2 = |V_{\PQtpr\PQbpr}|^2$. Degenerate masses have been assumed for the fourth-generation quarks, hence $m_{\PQtpr}=m_{\PQbpr}$. The information
is combined from different subsamples corresponding to different final states with at least one electron or muon. Observables
have been constructed in each of the subsamples and used to differentiate between the standard model background and the processes with
fourth-generation quarks. With this strategy the search for singly and pair-produced \PQtpr and \PQbpr quarks has been combined in a coherent way into a single analysis.
Model-dependent limits are derived on the mass of the quarks and the $V_{\mathrm{CKM}}^{4\times4}$ matrix element $A$. The existence of fourth-generation quarks with
masses below 685\GeV is excluded at 95\% confidence level for minimal off-diagonal mixing between the third- and the fourth-generation quarks. A non-zero cross section for
the single fourth-generation quark production processes, corresponding to a value of the $V_{\mathrm{CKM}}^{4\times4}$ parameter $A < 1$ gives rise to a more stringent
limit. When a mass difference of 25\GeV is assumed between \PQtpr
and \PQbpr quarks, the limit on $m_{\PQtpr}$ shifts by about $+20$ ($-20$)\GeV for $m_{\PQtpr}=m_{\PQbpr}+25\GeV$ ($m_{\PQtpr}=m_{\PQbpr}-25\GeV$).
These results significantly reduce the allowed parameter space for a fourth generation of fermions and raise the lower limits on the masses of the fourth generation quarks to the region where
nonperturbative effects of the weak interactions are important.

\section*{Acknowledgements}

We congratulate our colleagues in the CERN accelerator departments for the excellent performance of the LHC machine. We thank the technical and administrative staff at CERN and other CMS institutes, and acknowledge support from BMWF and FWF (Austria); FNRS and FWO (Belgium); CNPq, CAPES, FAPERJ, and FAPESP (Brazil); MES (Bulgaria); CERN; CAS, MoST, and NSFC (China); COLCIENCIAS (Colombia); MSES (Croatia); RPF (Cyprus); MoER, SF0690030s09 and ERDF (Estonia); Academy of Finland, MEC, and HIP (Finland); CEA and CNRS/IN2P3 (France); BMBF, DFG, and HGF (Germany); GSRT (Greece); OTKA and NKTH (Hungary); DAE and DST (India); IPM (Iran); SFI (Ireland); INFN (Italy); NRF and WCU (Korea); LAS (Lithuania); CINVESTAV, CONACYT, SEP, and UASLP-FAI (Mexico); MSI (New Zealand); PAEC (Pakistan); MSHE and NSC (Poland); FCT (Portugal); JINR (Armenia, Belarus, Georgia, Ukraine, Uzbekistan); MON, RosAtom, RAS and RFBR (Russia); MSTD (Serbia); SEIDI and CPAN (Spain); Swiss Funding Agencies (Switzerland); NSC (Taipei); ThEP, IPST and NECTEC (Thailand); TUBITAK and TAEK (Turkey); NASU (Ukraine); STFC (United Kingdom); DOE and NSF (USA).
Individuals have received support from the Marie-Curie programme and the European Research Council (European Union); the Leventis Foundation; the A. P. Sloan Foundation; the Alexander von Humboldt Foundation; the Austrian Science Fund (FWF); the Belgian Federal Science Policy Office; the Fonds pour la Formation \`a la Recherche dans l'Industrie et dans l'Agriculture (FRIA-Belgium); the Agentschap voor Innovatie door Wetenschap en Technologie (IWT-Belgium); the Ministry of Education, Youth and Sports (MEYS) of Czech Republic; the Council of Science and Industrial Research, India; the Compagnia di San Paolo (Torino); and the HOMING PLUS programme of Foundation for Polish Science, cofinanced from European Union, Regional Development Fund.

\bibliography{auto_generated}   

\cleardoublepage \appendix\section{The CMS Collaboration \label{app:collab}}\begin{sloppypar}\hyphenpenalty=5000\widowpenalty=500\clubpenalty=5000\input{EXO-11-098-authorlist.tex}\end{sloppypar}
\end{document}

%% file: EXO-11-098-authorlist.tex
\textbf{Yerevan Physics Institute,  Yerevan,  Armenia}\\*[0pt]
S.~Chatrchyan, V.~Khachatryan, A.M.~Sirunyan, A.~Tumasyan
\vskip\cmsinstskip
\textbf{Institut f\"{u}r Hochenergiephysik der OeAW,  Wien,  Austria}\\*[0pt]
W.~Adam, E.~Aguilo, T.~Bergauer, M.~Dragicevic, J.~Er\"{o}, C.~Fabjan\cmsAuthorMark{1}, M.~Friedl, R.~Fr\"{u}hwirth\cmsAuthorMark{1}, V.M.~Ghete, J.~Hammer, N.~H\"{o}rmann, J.~Hrubec, M.~Jeitler\cmsAuthorMark{1}, W.~Kiesenhofer, V.~Kn\"{u}nz, M.~Krammer\cmsAuthorMark{1}, I.~Kr\"{a}tschmer, D.~Liko, I.~Mikulec, M.~Pernicka$^{\textrm{\dag}}$, B.~Rahbaran, C.~Rohringer, H.~Rohringer, R.~Sch\"{o}fbeck, J.~Strauss, A.~Taurok, W.~Waltenberger, G.~Walzel, E.~Widl, C.-E.~Wulz\cmsAuthorMark{1}
% ...
\vskip\cmsinstskip
\dag:~Deceased\\
1:~~Also at Vienna University of Technology, Vienna, Austria\\
% ...
